\documentstyle[pre,aps,floats,epsf,amsmath,amssymb,psfig,epsfig]{revtex}

\newcommand{\PR}{{\em{Phys. Rev.~}}}
\newcommand{\JPA}{{\em{J. Phys. A}}}

\begin{document}

\title{Density waves and jamming transition in cellular automaton
  models for traffic flow}

\author{L Neubert\dag, H Y Lee\ddag\ and M Schreckenberg\dag}

\address{\dag\ FB 10, Theoretische Physik,
  Gerhard-Mercator-Universit\"at Duisburg, 47048~Duisburg, Germany,\\
  e-mail: \{neubert,schreck\}@traffic.uni-duisburg.de}

\address{\ddag\ Department of Physics and Centre for Theoretical Physics,
  Seoul National University, Seoul 151-742, Korea,\\ e-mail:
  agnes@phya.snu.ac.kr}

\maketitle

\begin{abstract}
  In this paper computer simulation results of higher order density
  correlation for cellular automaton models of traffic flow are
  presented. The examinations show the jamming transition as a
  function of both the density and the magnitude of noise and allow to
  calculate the velocity of upstream moving jams. This velocity is
  independent of the density and decreases with growing noise. The
  point of maximum flow in the fundamental diagram determines its
  value. For that it is not necessary to define explicitly jams in the
  language of the selected model, but only based upon the well defined
  characteristic density profiles along the line.
\end{abstract}

\section{Introduction}
\label{sec:intro}

Recently, the examination and modeling of vehicular traffic has become
an important subject of research -- see
\cite{tgf}-\nocite{Lighthill55a,Kerner93,Bando95b}\cite{Helbing97b}
and references therein for a brief review. In the microscopic approach
to the traffic flow problem, the cellular automaton introduced in
\cite{Nagel92} reproduces important entities of real traffic, like the
flow-density relation or stop-and-go waves. Beside the realization of
some basic requirements to such a model it can be efficiently used in
computational investigations and applications
\cite{Chopard96}-\nocite{Rickert96b,Esser97a}\cite{isttt}. Fundamental
analytical and numerical examinations enclose exact solutions for
certain limits and mean-field approximations \cite{Schadschneider},
the jamming transition
\cite{Csanyi95}-\nocite{Sasvari97,Eisen98,Luebeck98a,Cheybani98}\cite{Roters99}
or the effects of perturbations and the occurrence of meta-stable
states \cite{Krauss97}-\nocite{Schadschneider97c}\cite{Barlovic98a},
e.g. We investigate the density waves and the separation in free-flow
and dense regions by means of the density-autocorrelation function. It
enables us to trace back the spatio-temporal evolution of jams which
are stable during the measurement time, on condition that jams emerge.
It should be noted that the probability for a jam to survive decreases
with the simulation time \cite{Nagel94a} in a system without a clear
phase separation between congestion and free-flow. But the duration of
a simulation is sufficiently shorter than these time periods. By this
method it is superfluous to give an explicit definition of what a jam
is and which cars are belonging to the jam. Therefore this method can
be theoretically used for every traffic flow model where density
profiles are available. As an example we apply this method to cellular
automaton models. In this context we will report and discuss several
aspects of the underlying model and their slow-to-start modifications
(Section~\ref{sec:model}): The jamming transition shows up by varying
both the global density $\rho$ and the global noise $p$; the jam
velocity can be derived directly from the density-autocorrelation
function and is closely related to the global flow-density relation
(Section~\ref{sec:results}). It is not the goal of the paper to
discuss the jamming transition with regard to the criticality or the
sharpness of this issue, but we apply the method of correlation
function from a more practical point of view.

\section{The model}
\label{sec:model}
Within the framework of this paper we only consider a one-dimensional
ring of cells. The cells are either vacant or occupied by a vehicle
labeled $i$. Its position is $x_i$ and its discrete velocity is
$v_i\in [0,v_{max}]$. The gap $g_i$ denotes the number of empty sites
to its leading vehicle. The rules for a parallel update are
\begin{itemize}
\item Acceleration with regard to the vehicle ahead: $v_i' \leftarrow \min
  (v_i+1,g_i,v_{max})$,
\item Noise: with a probability $p$ do $v_i'' \leftarrow
  \max (v_i'-1,0)$,
\item Movement: $x_i \leftarrow x_i+v_i''$.
\end{itemize}
The investigated systems consist of $L$ cells and $N$ vehicles, the
global density is $\rho=N/L$. The flow is defined as $J=\langle
v\rangle\rho$ with the mean velocity $\langle v\rangle=\sum v_i/N$. In
the following the Nagel-Schreckenberg cellular automaton model
\cite{Nagel92} defined through the above set of rules is denoted by
SCA.

We extend our studies on further modifications of the SCA, namely on
models with slow-to-start rules. For the model with velocity-depending
randomization (VDR) \cite{Schadschneider97c,Barlovic98a} we set
$\tilde p(v_i=0)=\mbox{Min(}p+p_{VDR},1$). That leads to a reduced
outflow from a jam. Note that $v_i$ is the velocity before the first
update step is performed. The other modified model under consideration
is the T$^2$-model introduced by Takayasu and Takayasu
\cite{Takayasu93}. Here the headway $g_i$ of a vehicle $i$ controls
the acceleration: standing vehicles with a headway $g_i=1$ only speed
up with a probability $1-\tilde p$ with $\tilde
p=\mbox{Min(}p+p_{T^2},1$), whereas for all other ones the rules are
unchanged. Unlike the SCA with similar parameters, both models exhibit
a different behavior in the vicinity of the point of maximum flow
($\rho_{max}\equiv\rho(J_{max}),J_{max}$). They are capable to
generate metastable states in the adiabatic approach (for details see
\cite{Barlovic98a}), i.e.~one finds two branches of $J(\rho)$ in a
small density interval. Additionally, for sufficient small $p$ one can
found a clear separation of of dense and free-flow regions in a
space-time plot. We summarize the effective deceleration probabilities
of the applied models:
\begin{eqnarray}
  \begin{array}{rrcl}
    \mbox{SCA:} & p & =  & const. \\
    \mbox{VDR:} & \tilde p & = & \left\{
      \begin{array}{ll}
          \mbox{Min} (p+p_{VDR},1) & v_i=0 \\
          p & \mbox{otherwise}
        \end{array}
      \right.\\
    \mbox{T}^2: & ~~~~\tilde p & = &  \left\{
      \begin{array}{ll}
          \mbox{Min} (p+p_{T^2},1)~~ & v_i=0 \wedge g_i=1 \\
          p &  \mbox{otherwise}
        \end{array}
      \right.\\
  \end{array}
  \label{eq:p_def}
\end{eqnarray}
Actually, other definitions of $\tilde p$ are conceivable, but, for
our purpose, we decided to use only the above notations. Primarily, it
was done for modeling moving vehicles with similar properties and to
scan the parameter space by varying only $p$. Both $p_{VDR}$ and
$p_{T^2}$ are of any value but fix.

\section{Simulation results and their discussion}
\label{sec:results}

\begin{figure}
  \begin{center}
    \epsfig{file=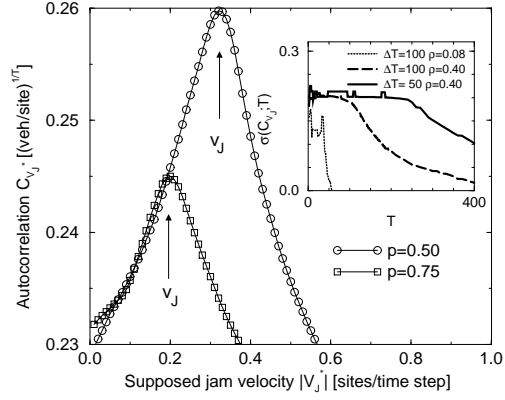,width=0.4\linewidth}
  \end{center}
 \caption{The peak of the density-autocorrelation function
    enables to estimate the jam velocity $V_J$.  The standard
    deviation of the symmetrically assumed $C_{V_J}$ is depicted in
    the inset. For large ratios $T/\Delta T$ the variance shrinks a
    lot, therefore it might happen that the autocorrelation function
    vanishes and inhibits the estimation of depending quantities. In
    the vicinity of $\rho^*$ this sensitivity is more pronounced
    ($\rho=0.4$).}
  \label{fig:autocorr}
\end{figure}

The density waves are moving upstream and can be easily observed in a
space-time plot \cite{Nagel92,Treiterer75}, one finds separation of
dense and free-flow regions. For the measurements it is necessary to
introduce the mean local density $\rho_l(k,t)$ of the cell $k$ at time
$t$:
\begin{eqnarray}
  \rho_l(k,t) = \frac{1}{\lambda}\sum_{i=0}^{\lambda-1}\eta_{k+i}(t)
  \label{eq:local_dens}
\end{eqnarray}
with
\begin{eqnarray}
  \eta_{k+i}(t) = \left\{ 
    \begin{array}{ll}
      1 & \mbox{if site~}k+i\mbox{~is occupied at time }t\\
      0 & \mbox{otherwise} \\
    \end{array}
  \right..\nonumber
\end{eqnarray}
The parameter $\lambda$ denotes the length of the interval on which
the local density has to be computed. It should satisfy the condition
$\lambda_0\ll\lambda\ll L$ \cite{Luebeck98a} with a characteristic
length scale $\lambda_0$. For the determination of the jam velocity
$V_J$ we use the generalized $T$-point-autocorrelation function of the
density
\begin{eqnarray}
  C_{V^*_J}(r\equiv V^*_J\tau\Delta T,\tau) = \langle \prod_{\tau =0}^{T-1}
  \rho_l(x+V^*_J\tau\Delta T,t+\tau\Delta T) \rangle_L
  \label{eq:corrfunct}
\end{eqnarray}
with the supposed jam velocity $V^*_J\in[-1,0]$. By varying $V^*_J$
one finds a largest $C_{V^*_J}(r,\tau)$ (Fig.~\ref{fig:autocorr}).
$\Delta T$ is the time interval between two single measurements which
contributes to (\ref{eq:corrfunct}).  Sufficiently large values of
$\Delta T$ are necessary to observe a macroscopic motion and to
determine $V_J$ with an adequate accuracy.  Unless otherwise
mentioned, we set $L=10^4$ and $\Delta T = 10^2$ in order to exclude
any finite size effects. Usually, we are averaging over $20$
simulation runs with $v_{max}=5$ and $\lambda=30$.

As pointed out in Fig.~\ref{fig:autocorr} one has to adjust thoroughly
the parameters $T$ and $\Delta T$. If $T$ is of the order of magnitude
of $\Delta T$ then the uncertainty of the measurement covers the
interesting signal and $C_{V_J}$ vanishes. This problem becomes more
serious while approaching $\rho^*$. In this region the calculations
are complicated additionally due to large fluctuations of $C_{V_J}$
itself.

\begin{figure}
  \begin{center}
    \epsfig{file=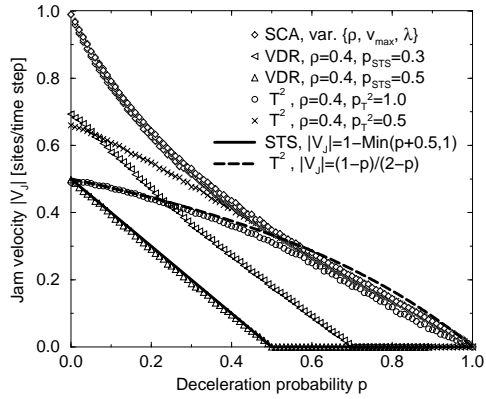,width=0.4\linewidth}
  \end{center}
  \caption{The jam velocity as a function of $p$, the error bars are
    within the symbol size ($\rho=0.4$).}
  \label{fig:vj_model}
\end{figure}

The jam velocity $V_J$ depends on the deceleration probability $p$
(Fig.~\ref{fig:vj_model}) and is related to global quantities as it
will be shown later. We checked this for a variety of parameters, but
could not notice any remarkable deviations among the diverse sets of
data. The modifications of the SCA, VDR and T$^2$, yield a different
behavior. The absolute values of the jam velocity in the models VDR
and T$^2$ are smaller than that of the SCA implying that both
modifications are characterized by a lowered outflow from a jam that,
in turn, reduces the jam velocity.

The results of simulations using the VDR are basically shifted towards
smaller $|V_J|$'s. $p_{VDR}$ can be recognized on the ordinate at
$p=0$ and on the abscissa at $V_J=0$, for large $p$'s a total deadlock
occurs, i.e.~it is highly unlikely or even impossible that a stopped
car speeds up again. In order to find out how $V_J$ is related to
$\tilde p$ we investigate the mean waiting time $t_w$ for $\tilde p<1$
expressed through an infinite series:
\begin{eqnarray}
   \nonumber
   t_w = 1(1-\tilde p) + 2(1-\tilde p)\tilde p + 3(1-\tilde p)\tilde
  p^2\cdots = (1-\tilde p)\sum_{n=1}^\infty n\tilde p^{n-1} =
   \frac{1}{1-\tilde p}\\
  \Rightarrow |V_J|=\frac{1}{t_w}=1-\tilde p~\in~[0, 1-p_{VDR}]
  \label{eq:vj_vdr}
\end{eqnarray}
and is exact for $p=0$. For large values of $p_{VDR}$ one obtains a
good agreement, whereas for small $p_{VDR}$ the jam velocity is
overestimated. This is due to the so-called sub-jams which emerge
downstream a wider jams and cause a reduction of $|V_J|$.

The results drawn from simulations using the T$^2$ model show no
deadlock situation for any $p<1$. Starting with $p=1$, one can hardly
distinguish between the simulation results of SCA and T$^2$.
Especially, this is valid as long as $p_{T^2}>1-p$.  Similar to
(\ref{eq:vj_vdr}) one can estimate
\begin{eqnarray}
  \nonumber t_w = 1(1-\tilde p) + 2(1-p)\tilde p + 3(1-p)\tilde pp +
  4(1-p)\tilde pp^2\cdots \\
  \phantom{t_w }= 1+\tilde p\sum_{n=0}^\infty p^n=
  \frac{1-p+\tilde p}{1-p}.
 \label{eq:vj_tt_ansatz}
\end{eqnarray}
Note that for small $p_{T^2}$-values $V_J$ is overestimated for all
values of $p$, which, in turn, can be traced back to the occurrence of
sub-jams. With increasing $p_{T^2}$ even for small $p$ it is required
to set $\tilde p =1$. Again, it is $|V_J|=t_w^{-1}$ and two special
cases can be described by
\begin{eqnarray}
  |V_J|(p=0)=\frac{1}{1+p_{T^2}} \qquad\mbox{and}\qquad 
  |V_J|(p_{T^2}\rightarrow 1)=\frac{1-p}{2-p}.
 \label{eq:vj_tt}
\end{eqnarray}
\begin{figure}
  \begin{center}
    \epsfig{file=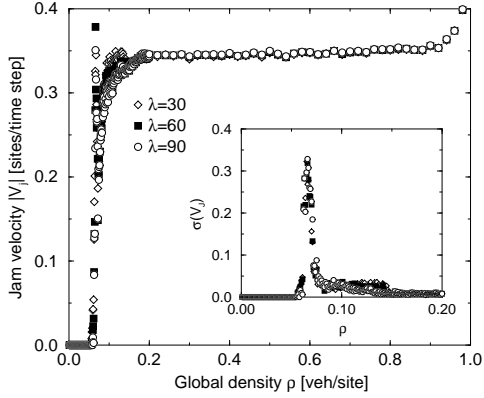,width=0.4\linewidth}
  \end{center}
  \caption{$V_J$ vs.~$\rho$ for $p=0.5$. Beyond $\rho^*$,
    especially for $0.2\leq\rho\leq 0.4$, $V_J(\rho)$ can be assumed
    to be constant. The inset reveals the large fluctuations of $V_J$
    nearby $\rho^*$ which are up to the order of magnitude of $V_J$
    itself.}
  \label{fig:vj_rho} 
\end{figure}
Obviously, the measurements reveals several density regimes. Below
$\rho^*$ the vehicles move independently, i.e.~there are no
correlations between them. For $\rho\geq\rho^*$ upstream moving
density waves can be detected by means of (\ref{eq:corrfunct}). In the
vicinity of $\rho^*$ the jam velocity reveals large fluctuations
(Fig.~\ref{fig:vj_rho}), which are due to the recurrent emergence and
dissipation of jams. But beyond $\rho^*$ $V_J$ is nearly constant.
Within the interval where density waves are to be expected it is
obvious that $V_J$ is independent of $\rho$, since the outflow from a
jam is independent from the global density.

\begin{figure}
  \begin{center}
    \epsfig{file=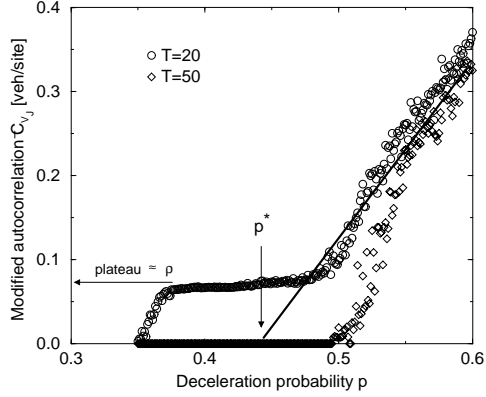,width=0.4\linewidth}
  \end{center}
  \caption{The transition from free flow to congested
    flow can also be obtained in the behavior of the modified
    autocorrelation function $\bar{C}_{V_J}$ (\ref{eq:modautocorr}) by
    varying $p$ ($\rho=0.073$). The transition is smeared out due to
    finite size effects and systematic errors in the determination of
    $\bar{C}_{V_J}$.}
  \label{fig:corr_p} 
\end{figure}

\begin{figure}
  \begin{center}
    \epsfig{file=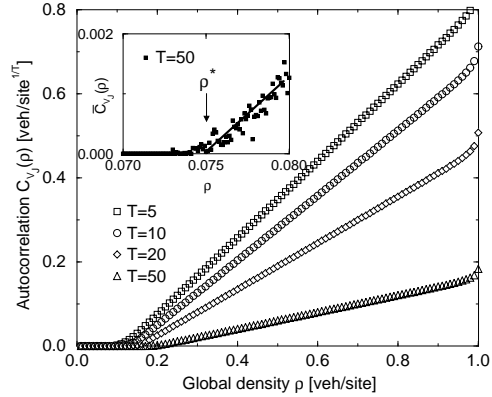,width=0.4\linewidth}
  \end{center}
  \caption{Plot of the the autocorrelation function
    $C_{V_J}(\rho)|_{v_{max}}$. Below a signified density the
    autocorrelation function vanishes due to the absence of jams. The
    inset zooms into the region $\rho\approx\rho^*$ for the modified
    autocorrelation $\bar{C}_{V_J}$ (\ref{eq:modautocorr}) ($p=0.5$).}
  \label{fig:corr_rho}
\end{figure}

\begin{figure}
  \begin{center}
    \epsfig{file=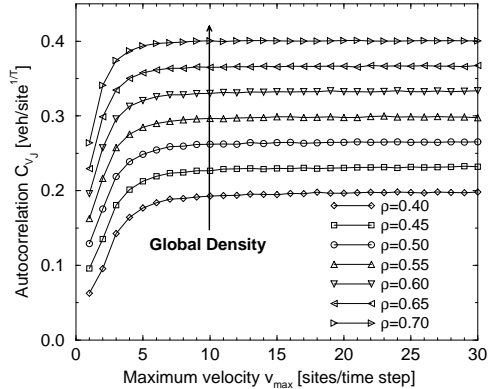,width=0.4\linewidth}
  \end{center}
  \caption{Plot of the the autocorrelation function
    $C_{V_J}(v_{max})|_{\rho}$ ($p=0.5$).}
  \label{fig:corr_vmax}
\end{figure}

So far, we applied the autocorrelation function (\ref{eq:corrfunct})
to determine the jam velocity. But this quantity itself indicates the
two different phases separated by noise $p^*$ or density $\rho^*$
(Fig.~\ref{fig:corr_p} and \ref{fig:corr_rho}). Varying $p$ leads to a
transition while crossing $p^*$. Its clarity strongly depends on
$T/\Delta T$, for insufficient ratios a plateau at $\bar{C}_{V_J}(p)$
occurs. To elucidate it we used a modified autocorrelation
\begin{eqnarray}
  \bar{C}_{V^*_J}(r,\tau) = \langle ( \prod_{\tau =0}^{T-1}
  \rho_l(x+V^*_J\tau\Delta T,t+\tau\Delta T) )^{1/T} \rangle_L.
  \label{eq:modautocorr}
\end{eqnarray}
The other transition takes place while crossing the density $\rho^*$
(Fig.~\ref{fig:corr_rho}). For $\rho<\rho^*$ one finds empty regions
on the road of the order of magnitude of $\lambda$, and therefore
$C_{V_J}$ completely vanishes. On the other hand, for $\rho>\rho^*$
stable congestion emerge. The same jam can be detected at $t_i$ as
well as at $t_f=t_i+\tau \Delta T$ located at $x(t_i)-|V_J|t_f$. In
this context, $\rho^*$ can be denoted as the density, at which stable
jams emerge and separates the density regime as it is to be seen in
the inset of Fig.~\ref{fig:corr_rho}. For a fixed density and a
varying $v_{max}$ (Fig.~\ref{fig:corr_vmax}) the relationship can be
estimated as $C_{V_J}(v_{max}) \propto \rho$. Nevertheless, the
quality of these data does not allow a correct classification of the
transition between the free-flow and the dense region. Above all, the
sensitive dependencies on the many adjustable parameters of this
method seem to prevent a more accurate consideration of the
interesting interval of density.

\begin{figure}
  \begin{center}
    \epsfig{file=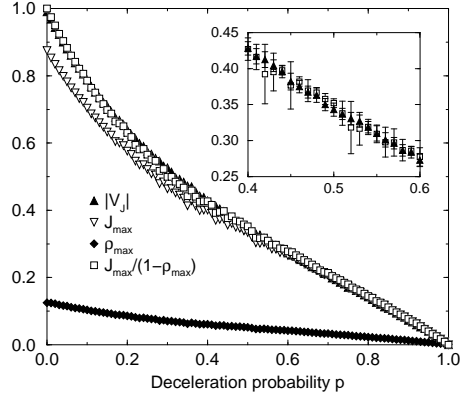,width=0.4\linewidth}
  \end{center}
  \caption{The jam velocity can be explained by (\ref{eq:vj_expl}):
    $V_J=J_{max}/(\rho_{max}-1)$.}
  \label{fig:expl}
\end{figure}

\begin{figure}
  \begin{center}
    \epsfig{file=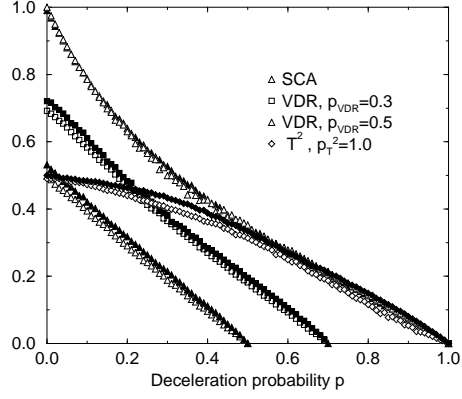,width=0.4\linewidth}
  \end{center}
  \caption{$V_J$ (filled symbols) as well as $J_{max}/(\rho_{max}-1)$
    (opaque symbols) are depicted for all used models. The small
    deviations are related to the discrepancies between the outflow of
    a jam and the global maximum flow.}
  \label{fig:expl_all}
\end{figure}

How is the jam velocity related to other macroscopic quantities? In
the steady state the dynamics are characterized by an equilibrium of
out-flowing vehicles and vehicles attaching the jam from behind. The
more frequent vehicles join the jam the faster the jam moves upstream.
If we neglect any effects due to meta-stability then the free-flow
region can be assumed to be located in the vicinity of the point of
maximum flow ($\rho_{max},J_{max}$). Hence, the velocity of attaching
vehicles is $\langle v_{att} \rangle =J_{max}/\rho_{max}$. The mean
distance between the upstream tail of the jam and the next vehicle is
$\langle g \rangle = \rho_{max}^{-1}-1$ and the temporal distance
therefore reads
\begin{eqnarray}
  \Delta t_{att}=\frac{\langle g \rangle}{\langle v_{att} \rangle}
  \quad \Leftrightarrow \quad 
  V_J = \frac{J_{max}}{\rho_{max}-1} \leq 0.
  \label{eq:vj_expl}
\end{eqnarray}
This is confirmed by the simulation results depicted in
Fig.~\ref{fig:expl}. It means that $V_J$ is determined by the slope of
the congested branch ($\rho\geq\rho_{max}$) in the fundamental
diagram.  This can also be verified for the modifications VDR and
T$^2$ (Fig.~\ref{fig:expl_all}). The small deviations from the data
rest upon a difference between the outflow of the jam and the maximum
global flow in the considered systems, but also in the above made
assumption of the equilibrium. Actually, the lowered outflow from jams
observed in the models VDR and T$^2$ in comparison to the SCA is also
reflected by (\ref{eq:vj_expl}).

Concluding, this knowledge enables a calibration of the SCA. Besides
the approach of the fundamental diagram derived from empirical data a
further point of interest is the velocity of upstream moving jams
($\approx -15~km/h$ on German highways \cite{Kerner96a}) -- but
according to (\ref{eq:vj_expl}) all information is accumulated in the
fundamental diagram, namely in the second characteristic slope of
$J(\rho)$. For the SCA one can set $v_{max}=5$ and $p=0.2\cdots 0.3$
to adapt the simulation to this empirical jam velocity.

\section{Summary}
\label{sec:sum}

We investigated the cellular automaton model for vehicular traffic in
order to get information about the density waves and their velocity.
Beside the standard Nagel-Schreckenberg cellular automata we also
included two slow-to-start modifications (VDR and T$^2$). Both
resemble the SCA except the rules for standing vehicles. Loosely
spoken, they result in a lower flow downstream a jam and a clear phase
separation for certain density regimes. For the determination of the
jam velocity we used the density-autocorrelation function
$C_{V_J}(r,\tau)$.  Despite the high computational efforts
(${\mathcal{O}}(L^2)$) this method was suitable to be applied for our
calculations. Moreover, a definition of jams is not necessary and
therefore the method can be applied to every model that provides
density profiles along the road.

The quantity $C_{V_J}(r,\tau)$ reflects the two different phases and
depends on the global density $\rho$. The density regime is separated
by $\rho^*$. For $\rho<\rho^*$ no jams can be detected by the applied
method, whereas for larger $\rho$ the system is dominated by sequences
of dense and free-flow regions, where $C_{V_J}$ remains finite and
permits to estimate $\rho^*$.  At this point a transition to the
congested region takes place. Both the local length $\lambda$ and the
number of calculations $T$ have large influence on $C_{V_J}$. Further
statements regarding the transition cannot be given due to the
numerical insufficiencies and accuracy.

The jam velocity can be derived directly from $C_{V_J}(r,\tau)$. For
sufficiently large $\rho$ the absolute value of $V_J$ is a continuous
and descending function of $p$, but depends on the considered model.
The differences between the models, especially for $p\rightarrow 0$
and $p\rightarrow 1$, could be explained by waiting time arguments.
The jam velocity is essentially expressed through $\rho_{max}$ and
$J_{max}$, irrespective of the model considered here.\\

{\bf Acknowledgement} The authors would like to thank A.
Schadschneider for helpful discussions and his important remarks on
this manuscript.

\end{document}